\documentclass[graybox]{svmult}

\usepackage{mathptmx}       
\usepackage{helvet}         
\usepackage{courier}        
\usepackage{type1cm}        
\usepackage{makeidx}         
\usepackage{graphicx}        
\usepackage{multicol}        
\usepackage[bottom]{footmisc}

\makeindex             

\begin{document}

\title*{Synthetic Cepheid Period-Luminosity \& Period-Color Relations in {\it Spitzer's} IRAC Bands}
\titlerunning{Synthetic IRAC Band P-L and P-C Relations} 
\author{C.-C. Ngeow, M. Marconi, I. Musella, M. Cignoni, M. Marengo \& S. Kanbur}
\authorrunning{Ngeow et al.} 
\institute{Chow-Choong Ngeow \at National Central University, Jhongli City, 32001, Taiwan. \email{cngeow@astro.ncu.edu.tw}
}
%
%
\maketitle

\vspace{-2.8cm}

\abstract*{The mid-infrared period-luminosity (P-L) relation for Cepheids will be important in the {\it JWST} era, as it holds the promise of deriving the Hubble constant within 2\% accuracy. We present the synthetic P-L and period-color (P-C) relations, derived from a series of stellar pulsation models with varying helium and metallicity abundance, in the {\it Spitzer} IRAC bands. Selected synthetic P-L and P-C relations were compared to the empirical relations derived from Galactic and Large Magellanic Cloud Cepheids. We also present the empirical P-L relations based on the latest distance measurements using infrared surface brightness techniques.}

\abstract{The mid-infrared period-luminosity (P-L) relation for Cepheids will be important in the {\it JWST} era, as it holds the promise of deriving the Hubble constant within 2\% accuracy. We present the synthetic P-L and period-color (P-C) relations, derived from a series of stellar pulsation models with varying helium and metallicity abundance, in the {\it Spitzer} IRAC bands. Selected synthetic P-L and P-C relations were compared to the empirical relations derived from Galactic and Large Magellanic Cloud Cepheids. We also present the empirical P-L relations based on the latest distance measurements using infrared surface brightness techniques.
}

\vspace{-0.6cm}
\section{Introduction}\label{sec:1}

The period-luminosity (P-L, also known as Leavitt Law) relation for classical Cepheid variables (hereafter Cepheids) is an important tool in astrophysics, as it is the first rung in distance scale ladder, and can be used to constraint stellar pulsation models. Investigations of Cepheid P-L relations, both empirically and theoretically, have been conducted mainly in optical and near infrared bands (for example, see \cite{bon10} and reference therein). Recently, researches on Cepheid P-L relations have been expanded to the mid-infrared (MIR), as MIR P-L relations will be important in the {\it James Webb Space Telescope (JWST)} era. Observations with {\it Spitzer} and/or {\it JWST} in MIR offer reductions of several key systematic errors in Hubble constant\footnote{For example, one of systematic errors being the extinction, as it is negligible in MIR.}. This opens the possibility of deriving the Hubble constant, via the cosmic distance ladder, to $\sim2\%$ level (\cite{fre10,fre11}). Several empirical studies of MIR P-L relations in {\it Spitzer's} IRAC bands have already been presented in \cite{fre08,nge08,mad09,nge09,mar10,nge10,sco11}. In terms of theoretical investigation, \cite{nge11} present the synthetic IRAC band P-L and period-color (P-C) relations constructed from state-of-the-art stellar pulsation codes. The aim of this paper is to provide an overview of these synthetic P-L and P-C relations, and compare them to the empirical P-L and P-C relations.

\vspace{-0.6cm}
\section{The Synthetic P-L and P-C Relations}\label{sec:2}

\begin{figure}
\includegraphics[scale=.3]{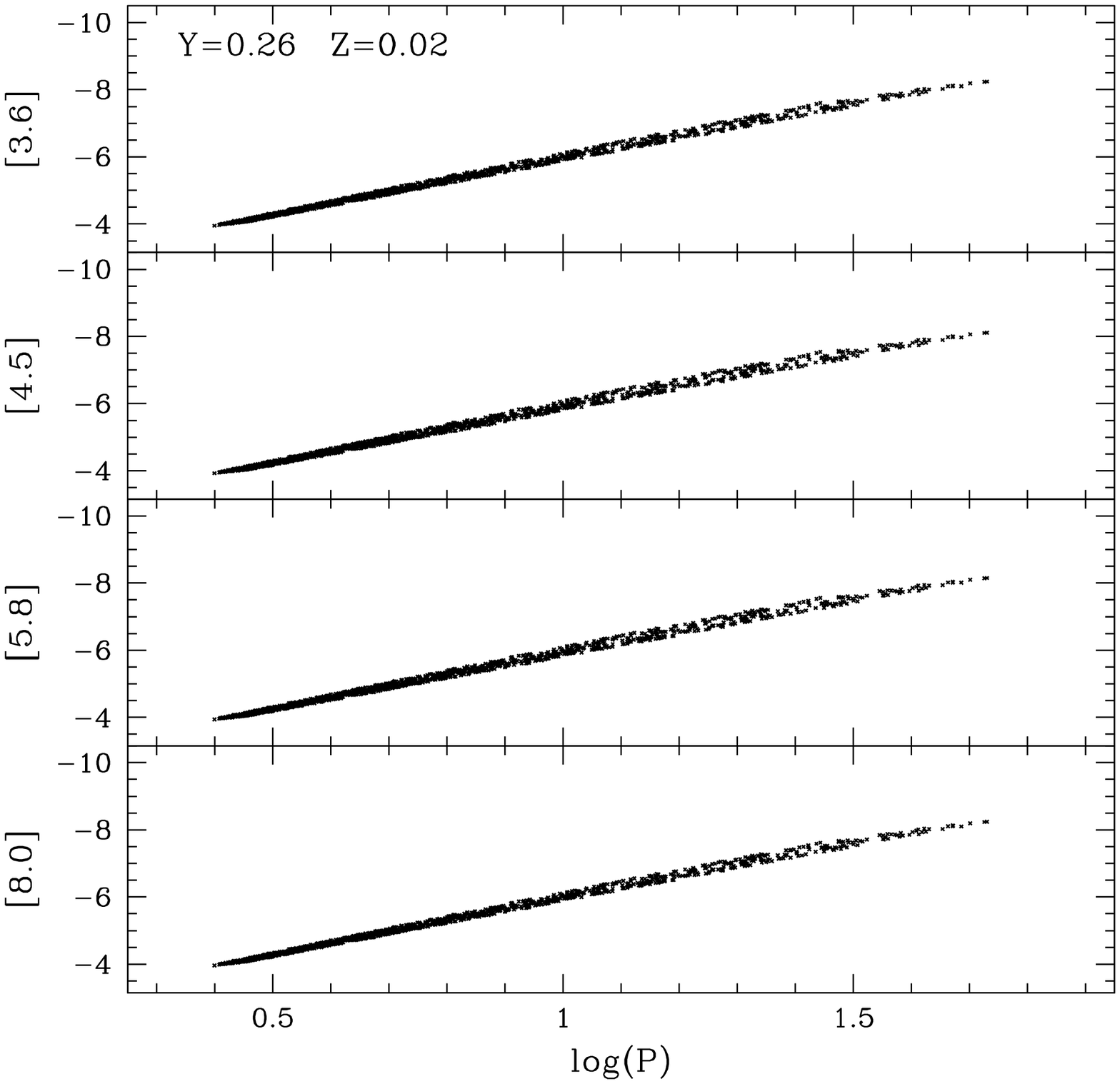}
\includegraphics[scale=.3]{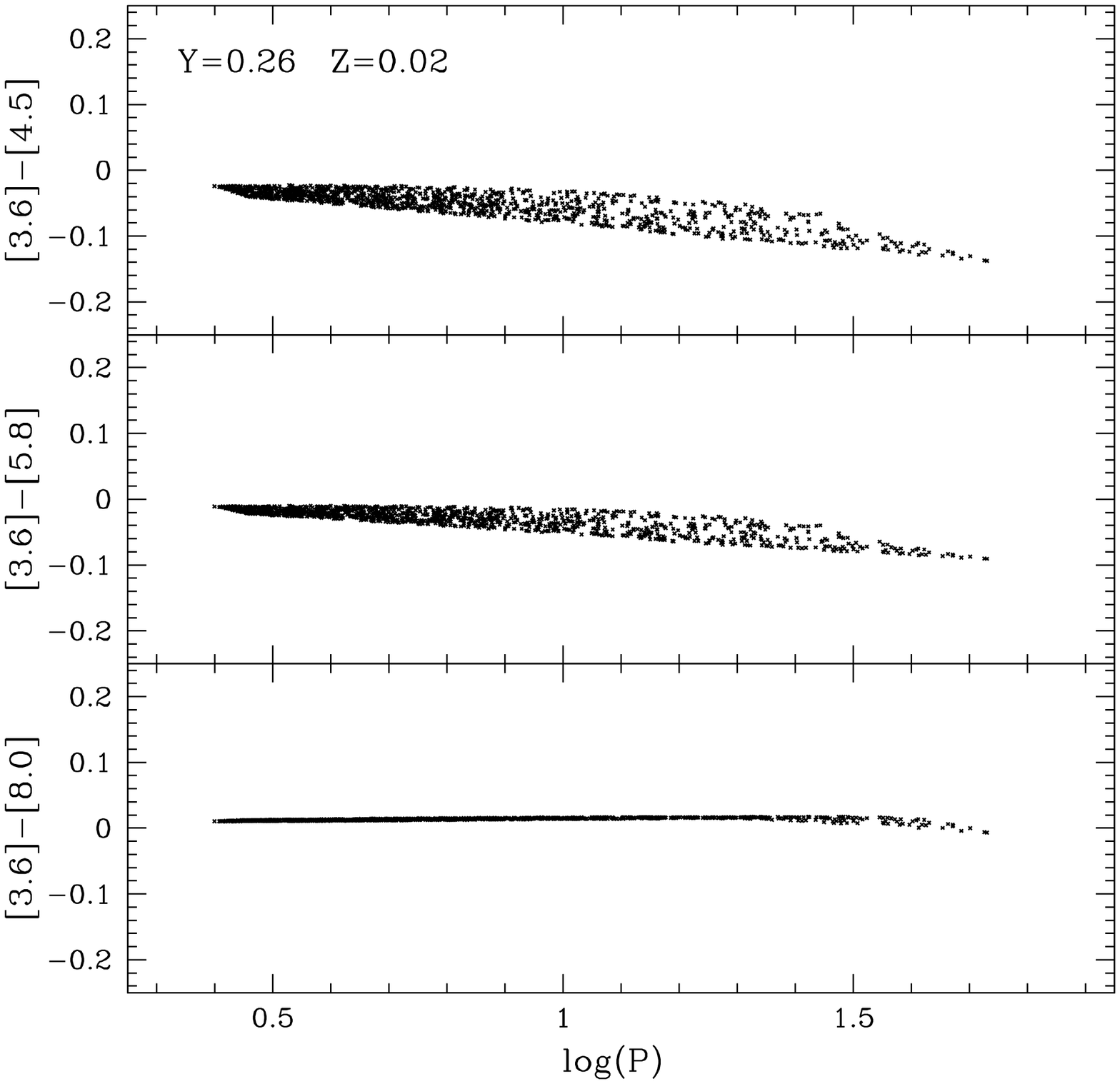}
\caption{Example of synthetic P-L relations (left panels) and P-C relations (right panels) based on $Y=0.26$, $Z=0.02$ model set.}
\label{fig_plpc}
\end{figure}

The synthetic P-L and P-C relations were generated from a series of stellar pulsation models using different inputs of metallicity ($Z$) and helium abundance ($Y$). Procedures and input physics for these stellar pulsation models are same as those used for theoretical investigation of optical and near infrared P-L relations, as presented in \cite{bon10}, with the exception that bolometric light curves were converted to IRAC band magnitudes using model atmospheres. In brief, the non-linear stellar pulsation models used in this work include a treatment of time-dependent coupling between pulsation and convection, and a canonical mass-luminosity relation. About $10^3$ pulsators generated from the pulsation models were used to populate the instability strip according to a mass law. Further details can be found in \cite{nge11} and reference therein. Figure \ref{fig_plpc} presents an example of the synthetic P-L and P-C relations from one of the model sets. 

\vspace{-0.6cm}
\section{Comparison to Observations}\label{sec:3}

\begin{figure}
\includegraphics[scale=.3]{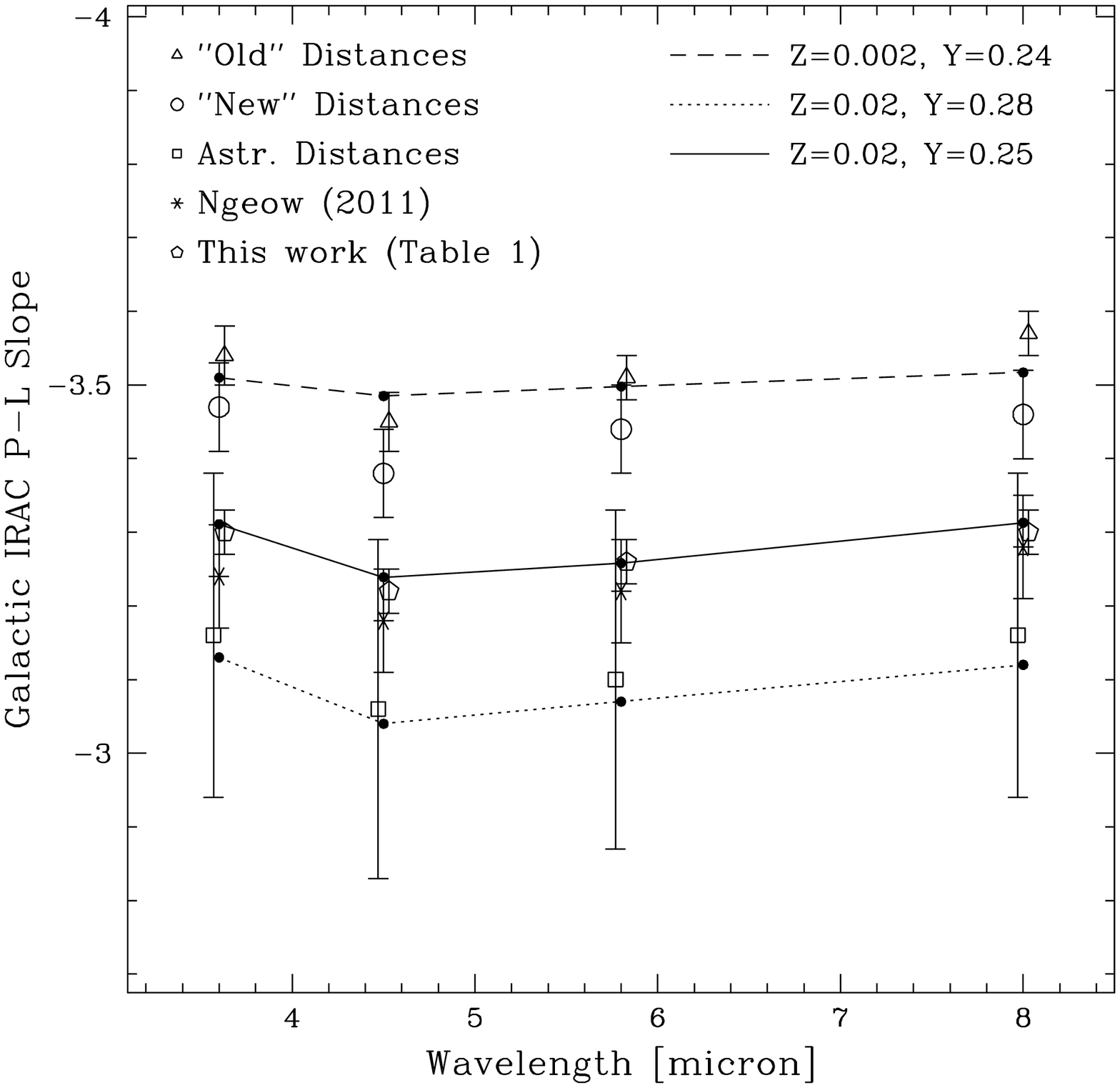}
\includegraphics[scale=.3]{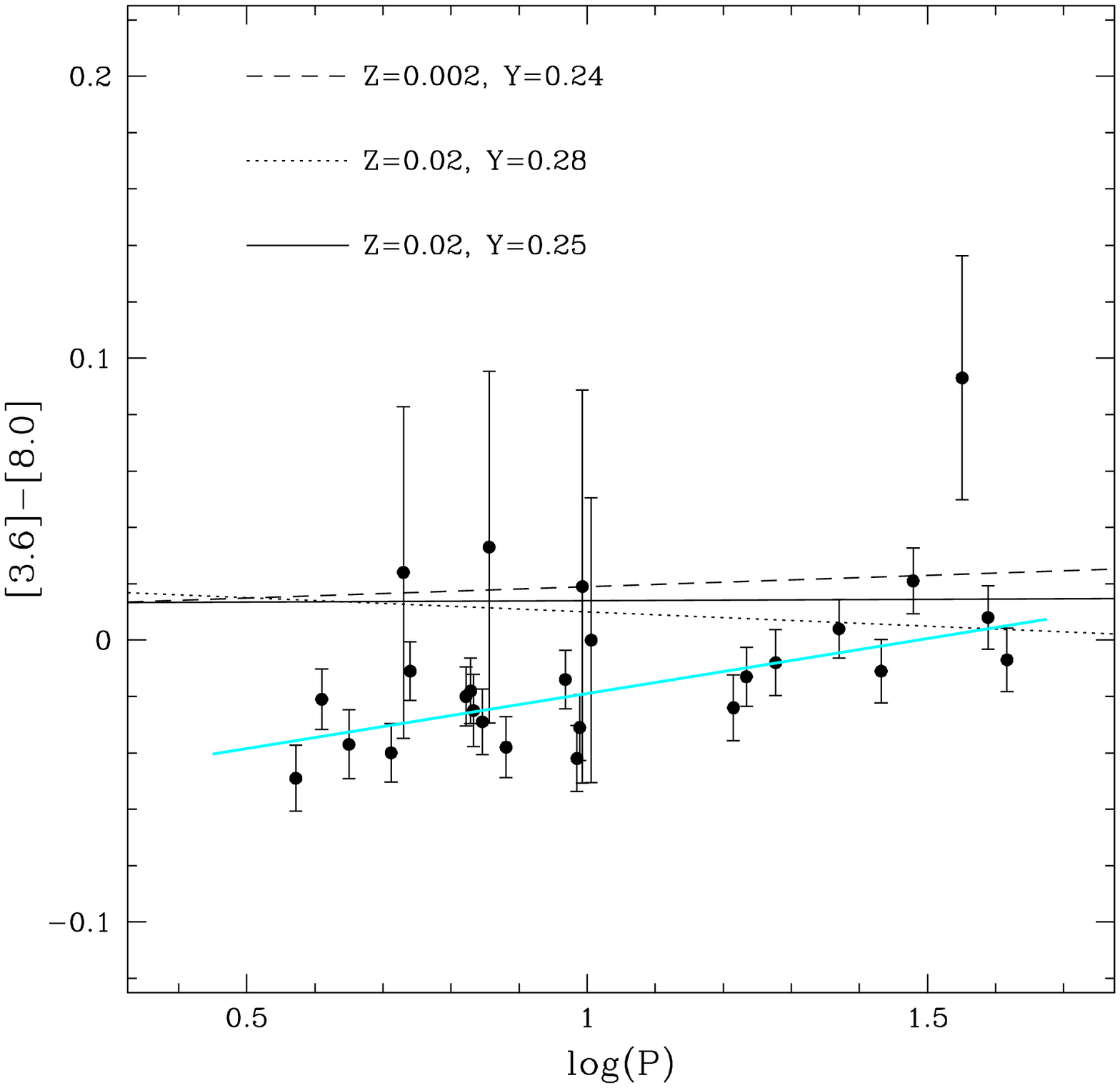}
\caption{{\bf Left Panel:} Comparison of the empirical Galactic P-L slopes to the synthetic P-L slopes from selected model sets. The P-L slopes based on ``Old'', ``New'' and Astr distances are adopted from \cite{mar10}. Wavelengths of the data points have been shifted slightly for better visualization. {\bf Right Panel:} Comparison of the synthetic P-C relations to the empirical Galactic P-C relation, where the data points are taken from \cite{mar10}. The cyan (or gray) line is the fitted P-C relation to the data (\cite{mar10}).}
\label{fig_gal}
\end{figure}

\cite{mar10} presented single-epoch {\it Spitzer} observations for about $29$ nearby Galactic Cepheids, and derived the empirical P-L relations for these Cepheids using three sets of distance measurements. These include the Baade-Wesselink type infrared surface brightness (IRSB) techniques with either an ``old'' or a ``new'' projection ($p$) factor, and geometrical parallaxes measurements. In addition, \cite{ngew} has also derived the IRAC band P-L relations for these Cepheids based on distance measurements using a calibrated Wesenheit function. Finally, the latest IRSB distances for these Cepheids are available from \cite{sto11a}, with a steeper $p$-factor relation. Hence, we re-derive the P-L relations using either the geometric parallaxes, if available, or these latest IRSB distances. The derived P-L relations are summarized in Table \ref{tab}. These five sets of empirical P-L slopes are compared to the selected synthetic P-L slopes in left panel of Figure \ref{fig_gal}. It is worth to point out that the empirical P-L slopes become shallower and closer to those based on geometrical parallaxes when the adopted $p$-factor relation is steeper in IRSB techniques. Right panel of Figure \ref{fig_gal} compares the synthetic P-C relations with empirical result, disagreement can be seen between synthetic and empirical P-C relations.

\vspace{-0.7cm}
\begin{table}
\caption{New empirical IRAC band P-L relations for Galactic and LMC Cepheids.}
\label{tab}       
\begin{tabular}{p{1.8cm}p{2.2cm}p{2.2cm}p{2.2cm}p{2.2cm}}
\hline\noalign{\smallskip}
Band: & $3.6\mu\mathrm{m}$ & $4.5\mu\mathrm{m}$ & $5.8\mu\mathrm{m}$ & $8.0\mu\mathrm{m}$ \\
\noalign{\smallskip}\svhline\noalign{\smallskip}
\multicolumn{5}{c}{Galactic Cepheid} \\
Slopes     & $-3.303\pm0.031$ & $-3.216\pm0.031$ & $-3.257\pm0.027$ & $-3.299\pm0.025$ \\
Intercepts & $-2.429\pm0.034$ & $-2.482\pm0.035$ & $-2.414\pm0.030$ & $-2.411\pm0.028$ \\
\multicolumn{5}{c}{LMC Cepheid} \\
Slopes     & $-3.268\pm0.042$ & $-3.142\pm0.042$ & $\cdots$ & $\cdots$ \\
Intercepts & $-2.544\pm0.061$ & $-2.682\pm0.061$ & $\cdots$ & $\cdots$ \\
\noalign{\smallskip}\hline\noalign{\smallskip}
\end{tabular}
\end{table}

\begin{figure}
\includegraphics[scale=.3]{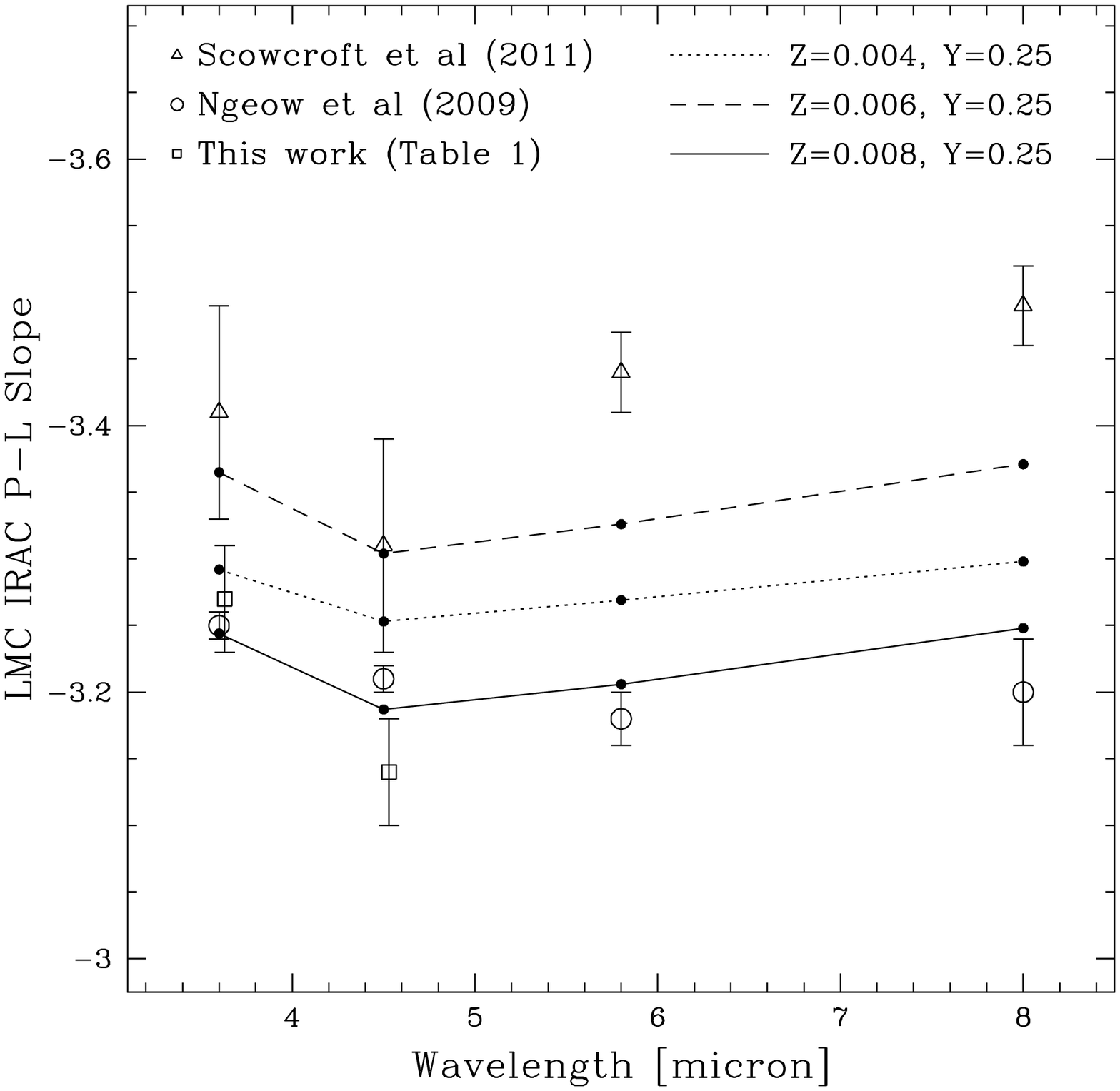}
\includegraphics[scale=.3]{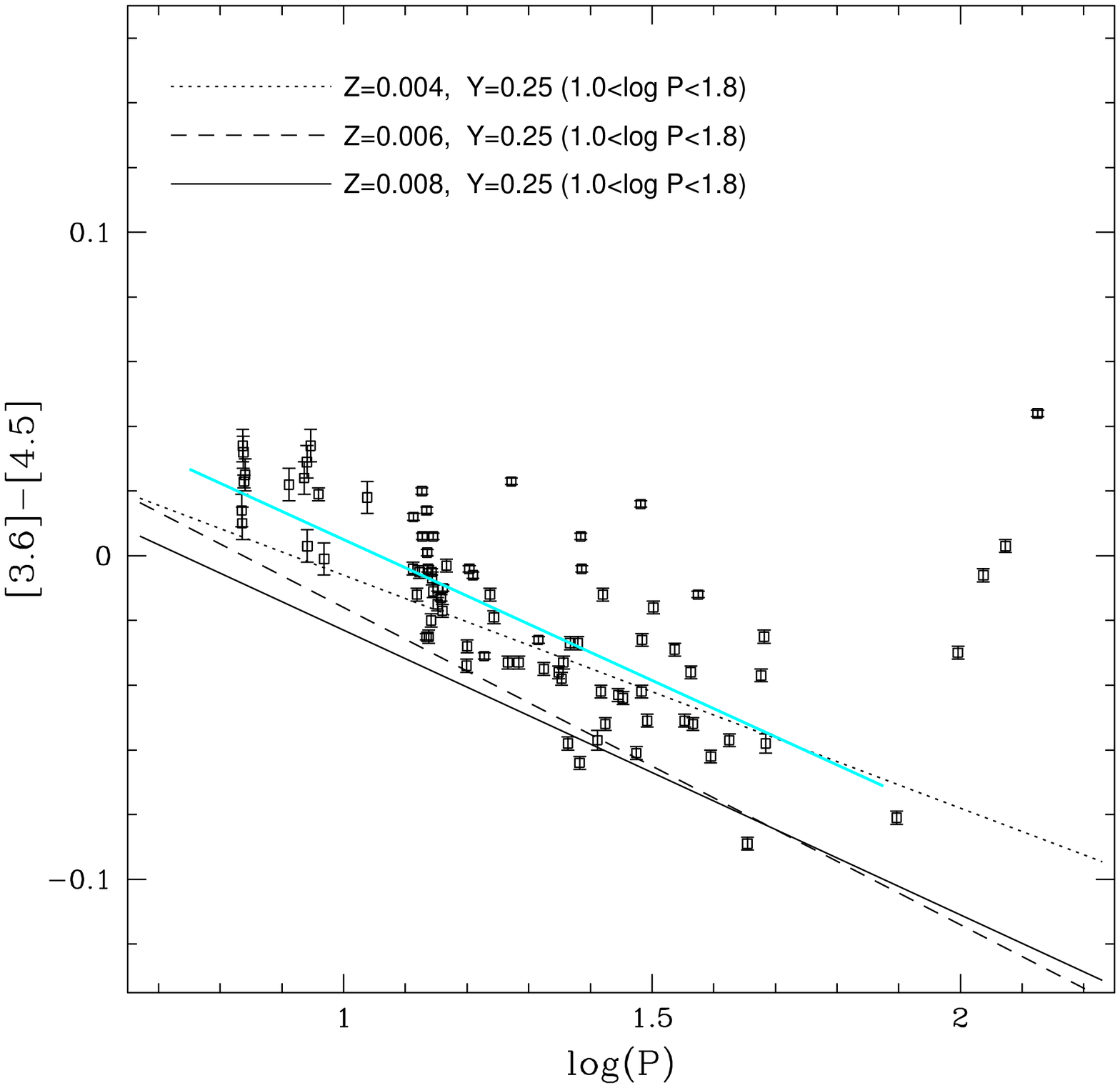}
\caption{{\bf Left Panel:} Comparison of the empirical LMC P-L slopes to the synthetic P-L slopes from selected model sets. Scowcroft et al. (2011) P-L slopes include those from \cite{sco11} (for $3.6\mu\mathrm{m}$ \& $4.5\mu\mathrm{m}$ band) and \cite{mad09} (for $5.8\mu\mathrm{m}$ \& $8.0\mu\mathrm{m}$ band). Wavelengths of the data points have been shifted slightly for better visualization. {\bf Right Panel:} Comparison of the synthetic P-C relations to the empirical LMC P-C relation, where the data points are taken from \cite{sco11}. The cyan (or gray) line is the fitted P-C relation to the data (\cite{sco11}).}
\label{fig_lmc}
\end{figure}

Empirical P-L relations for Large Magellanic Cloud (LMC) Cepheids were compared to the synthetic P-L slopes from selected models sets in left panel of Figure \ref{fig_lmc}. These empirical P-L slopes include those presented in \cite{nge09} and \cite{sco11,mad09}. Note that the $3.6\mu\mathrm{m}$ \& $4.5\mu\mathrm{m}$ band mean magnitudes given in \cite{sco11} are based on the available full light curves of $67$ LMC Cepheids, and $29$ of them have new IRSB distance measurements from \cite{sto11b}. The derived P-L relations by combining \cite{sco11} photometry and \cite{sto11b} distances, for Cepheids with period between $10$ and $60$ days, are summarized in Table \ref{tab} and compared to the synthetic P-L slopes in Figure \ref{fig_lmc}. Right panel of Figure \ref{fig_lmc} compares the selected synthetic P-C relations to the empirical P-C relation given in \cite{sco11}. See \cite{nge11} for further discussion of the comparison of these P-C relations.

\begin{acknowledgement}
CCN thanks the funding from National Science Council (of Taiwan) under the contract NSC 98-2112-M-008-013-MY3.
\end{acknowledgement}

\end{document}